\documentclass[aip,twocolumn,english]{revtex4}
\usepackage[T1]{fontenc}
\usepackage[latin9]{inputenc}
\usepackage{babel}
\usepackage{graphicx}

\begin{document}

\title{High frequency electro-optic measurement of strained silicon racetrack
resonators}

\author{M.Borghi, M.Mancinelli, L.Pavesi}
\affiliation{Department of Physics, Nanoscience Laboratory, University of Trento,
I-38123 Povo, Italy}
\author{F.Merget, J.Witzens}
\affiliation{Institute of Integrated Photonics, RWTH Aachen University, Sommerfeldstr.
24, D-52074, Aachen, Germany}
\author{M.Bernard, M.Ghulinyan, G.Pucker}
\affiliation{Centre for Materials and Microsystems, Fondazione Bruno Kessler,
I-38123 Povo, Italy}

\begin{abstract}
\noindent The observation of the electro-optic effect in strained silicon waveguides has been considered as a direct manifestation of an induced $\chi^{(2)}$ non-linearity in the material. In this work, we perform high frequency measurements on strained silicon racetrack resonators. Strain is controlled by a mechanical deformation of the waveguide. It is shown that any optical modulation vanishes independently of the applied strain when the applied voltage varies much faster than the carrier effective lifetime, and that the DC modulation is also largely independent of the applied strain. This demonstrates that plasma carrier dispersion is responsible for the observed electro-optic effect. After normalizing out free carrier effects, our results set an upper limit of $8\,pm/V$ to the induced high-speed $\chi^{(2)}_{\textup{eff},zzz}$ tensor element at an applied stress of $-0.5\,GPa$. This upper limit is about one order of magnitude lower than the previously reported values for static electro-optic measurements. 
\end{abstract}

\maketitle

\noindent In the last years, a lot of effort was spent investigating the strain induced second order nonlinearity ($\chi^{(2)}$ effect) in Silicon by investigating the induced Pockels effect \cite{jacobsen,chim,chim2,puckett,pedro}. Strain induced $\chi^{(2)}$ could be instrumental to make ultrafast and energy efficient electro-optic modulators for the Silicon photonics platform which would replace present electro-optic modulators based on the plasma dispersion effect \cite{reed,vivien,xu,Liu}. More generally, the presence of an appreciable $\chi^{(2)}$ in Silicon would validate the Silicon-on-Insulator (SOI) platform as an alternative to Lithium Niobate for second order nonlinear optics \cite{cazza,mher}. In most of these works, the centro-symmetry of Silicon is broken by a stressing film of Silicon Nitride which induces strain in the underlying Silicon waveguide and enables a $\chi^{(2)} \neq 0$ - at least from a first principle point of view \cite{Costanza}. With the exception of refs. \cite{cazza,mher}, the Pockels effect was investigated by using an integrated imbalanced Mach-Zehnder interferometer in which one or both interferometer arms are driven by a DC or a low frequency ($\approx kHz$) AC electric field \cite{jacobsen,chim,chim2,puckett,pedro}. Then, an effective $\chi^{(2)}_{\textup{mat}}$ value is extracted from the measured shift of the interference fringes by taking into account the system geometry and the magnitude of the applied static electric field. As expected from theory, the relation between the effective index change and the applied static electric field is found to be linear. The linear relation between these two physical quantities is considered as \emph{the} evidence of the observation of a Pockels effect. Unfortunately, a linear effective index variation of the optical mode of a waveguide can also be induced by free carriers \cite{azadeh,rajat} or by trap states and localized charges at the SiN$_x$-Si interface \cite{azadeh,mher}. A definitive proof of the strain induced non-linearity can be obtained by high frequency measurements in an interferometer structure since the Pockels effect and the free carrier dispersion are characterized by two different characteristic times. In this letter, we measure the separate contributions to the effective $\chi^{(2)}_{\textup{eff}}$ of the Pockels effect and of the plasma dispersion effect. We investigate the frequency response of the electro-optic effect in a racetrack resonator up to $5\,GHz$, i.e., well above the effective free carrier lifetime in the waveguide. In particular, by using a dedicated setup, we were able to tune continuously the strain in a same device so as to verify accurately how the applied strain modifies the system's frequency response. \\
\begin{figure*}[t!]
\centering
\includegraphics[scale=0.46]{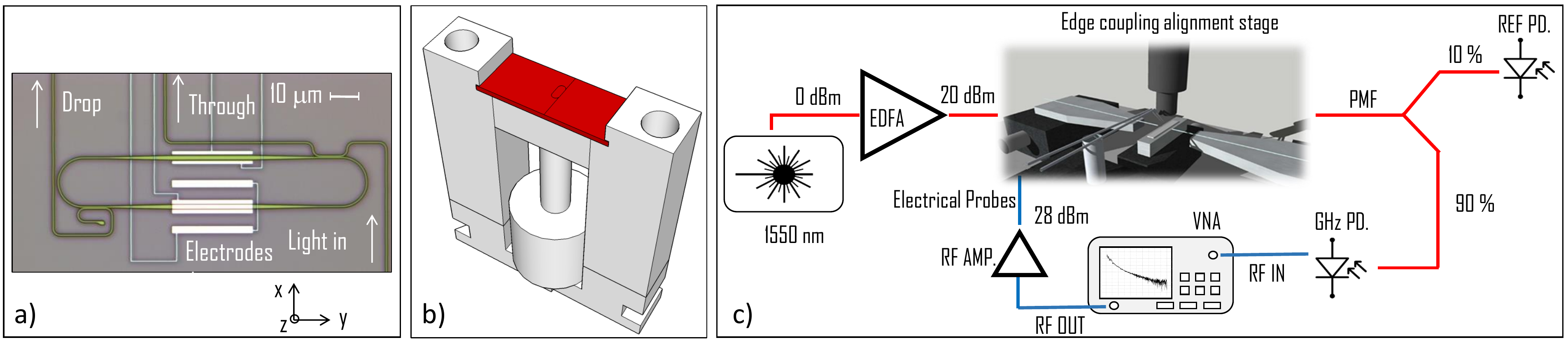}
\caption{(a) Optical microscope image of the racetrack resonator. Electrodes appear as white rectangles. Only the three electrodes on the bottom are used to apply an electric field in the $z$ direction. The narrow metallic stripes connect the electrodes to $150\,\mu m \times 50\, \mu m$ rectangular pads (not shown), which are used as contact points for the Tungsten tips. (b) 3D model of the stressing sample holder. The sample is indicated in red. (c) Experimental setup used for the electro-optic measurements. PMF = Polarization Maintaining Fibers, PD = Photodiodes, VNA = Vector Network Analyzer, RF IN(OUT) = Radio Frequency IN(OUT). }
\label{fig:figure1}
\end{figure*}
\begin{figure}[h!]
\centering
\includegraphics[scale=0.6]{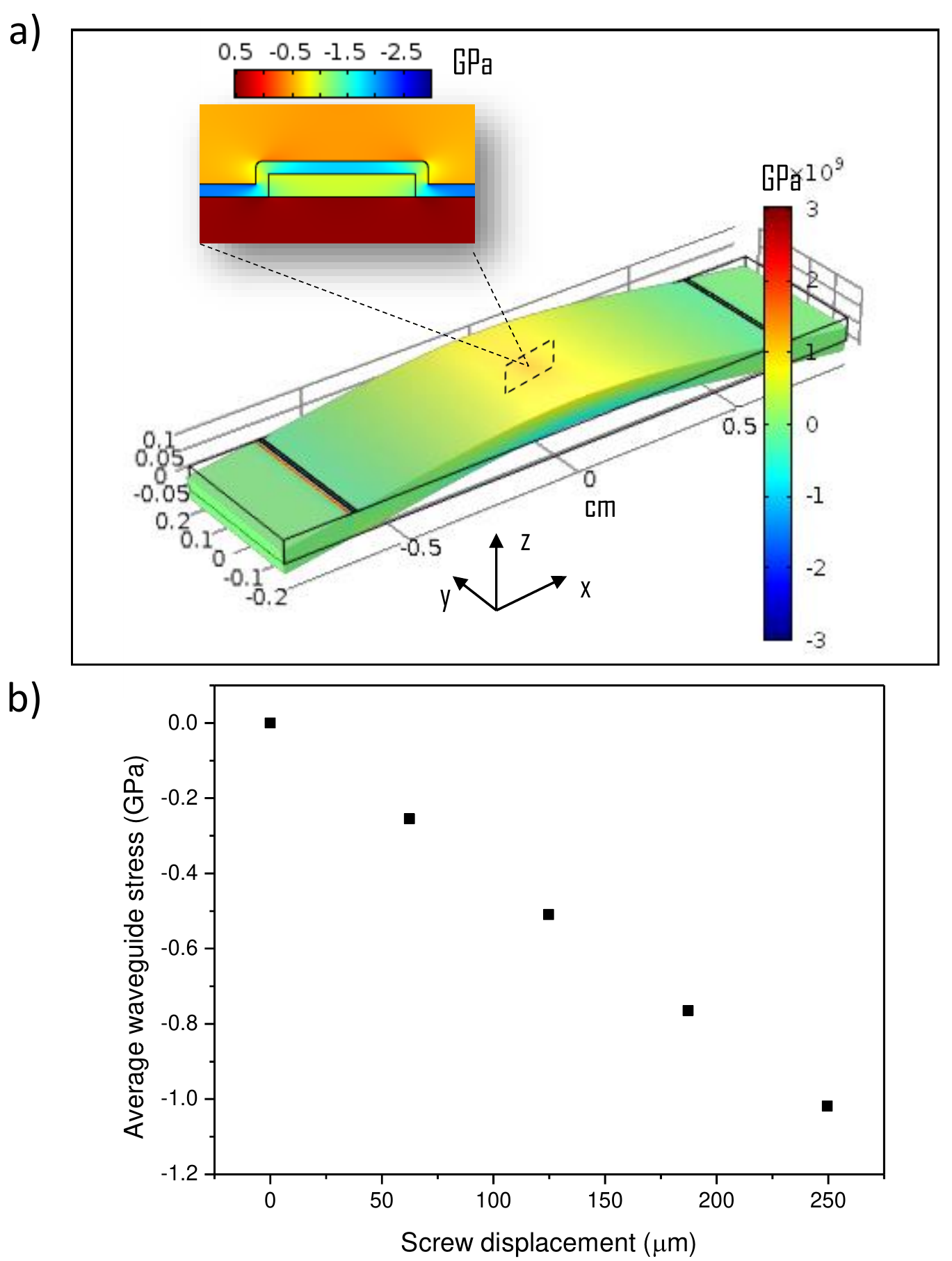}
\caption{(a) Finite Element simulation of the stress distribution ($\sigma_{xx}$ element of the stress tensor $\sigma_{ij}$) on the sample subjected to a $62.5\,\mu m$ screw displacement. The discontinuities near the ends are due to the line contact with the sample holder (see fig. \ref{fig:figure1}(b)), which lies $2\,mm$ from the end. The inset shows the stress profile along the waveguide cross section. (b) Average stress ($\sigma_{xx}$) inside the waveguide as a function of the screw displacement.}
\label{fig:figure2}
\end{figure}
Our test structure is shown in fig.\ref{fig:figure1}(a). It consists of a racetrack resonator in the Add-Drop filter configuration \cite{Masi} fabricated on a 6'' SOI wafer. The resonator has a perimeter of $415\,\mu m$ and a coupling coefficient with the bus waveguide of $\kappa^2=7\%$. A $140\,nm$ thick LPCVD Silicon Nitride ($Si_3N_4$) layer is conformally deposited on the silicon waveguide. We use a $3\,\mu m$ thick Buried Oxide Layer (BOX) and a $900\,nm$  thick Silica layer as lower and upper cladding materials. The residual stress on the $Si_3N_4$ layer has been measured to be $-0.19\,GPa$. An electric field can be applied in the \emph{z} direction (see fig.\ref{fig:figure1}(a)) using three Aluminum electrodes along a $50\,\mu m$ straight waveguide section. With reference to the geometry shown in fig.\ref{fig:figure1}(a), this is achieved by grounding the central electrode on the top of the waveguide, and by shorting the two adjacent electrodes to a common voltage. This configuration is similar to the one reported in ref. \cite{pedro}. The waveguide width changes adiabatically from $400\,nm$ outside the electrode region to $1600\,nm$ below them. The waveguide height is kept fixed at $250\,nm$. Using the adiabatic tapers, we are able to probe the electro-optic modulation in the $1600\,nm$ wide multimode waveguide by preserving the fundamental mode excitation. As shown in fig.\ref{fig:figure1}(b), the device is fixed on a sample holder which can provide a variable stress adjusted by rotating a $250\,\mu m$ pitch screw. The material strain is mechanically induced by fixing the sample ends to the holder and by subsequently bending its center by means of the pressure exercised by the screw. We found that it is possible to displace the screw by approximately its complete pitch before breaking the chip. To avoid micro fractures, we kept screw displacements lower than $150\mu m$. It is important to note that, with respect to other studies in which the analysis is performed by comparing samples with different strain \cite{jacobsen,chim,chim2,puckett,pedro,cazza,mher}, in our case it is possible to tune continuously the induced strain in the very same sample/resonator. As indicated in fig. \ref{fig:figure1}(c), light from a C-band infra-red laser amplified by an Erbium Doped Fiber Amplifier (EDFA) is edge coupled to the Input port of the resonator using a Polarization Maintaining Lensed Fiber (PMF). The light polarization is set to Transverse Magnetic (TM). A nanometric XYZ positioning stage is used to minimize the coupling losses. The transmitted light from the resonator Through port is split: $10\%$ is sent to a reference photodiode, while  $90\%$ feeds a high bandwidth photoreceiver ($43\,GHz$) connected to a Vector Network Analyzer (VNA). The VNA also provides $28\,dBm$ (after amplification) of sinusoidal voltage modulation to the electrodes using impedance matched Tungsten tips with a $40\,GHz$ bandwidth. \\
We determined the relation between the screw displacement and the stress induced inside the waveguide from Finite Element Method (FEM) computations. The results are shown in fig.\ref{fig:figure2}(a-b). Only the $\sigma_{xx}$ element of the stress tensor $\sigma_{ij}$ is plotted, since it is found to be one order of magnitude higher than the $\sigma_{yy}$ component and five orders of magnitude higher than the remaining tensor elements. In the region where the resonator is located (dashed rectangle in fig. \ref{fig:figure2}(a)), the overall stress distribution can be approximated as compressive and uniaxial in the $x$ direction. In fig.\ref{fig:figure2}(b), we show a linear relation between the screw displacement and the computed average stress in the waveguide. At our working point (displacement $\approx 125\,\mu m$), the stress level is about $-0.48\,GPa$ in the waveguide and $-0.78\, GPa$ in the $Si_3N_4$ layer. The stress magnitude and direction is comparable to the ones used in other experiments \cite{jacobsen,chim,chim2,puckett,pedro,cazza,mher}. In the inset of fig. \ref{fig:figure2}(a) we also notice that a high stress (hence strain) gradient is present at the upper and lower interfaces between $Si_3N_4$ and the waveguide. This feature is essential since it is theoretically predicted that the induced $\chi^{(2)}$ is proportional to the strain gradient in the material \cite{Costanza,cazza}. 
At first, we set the screw displacement to zero, so no stress (except the residual one due to the $Si_3N_4-Si$ interface) is applied to the waveguide. 

\begin{figure*}[t!]
\centering
\fbox{\includegraphics[scale=0.48]{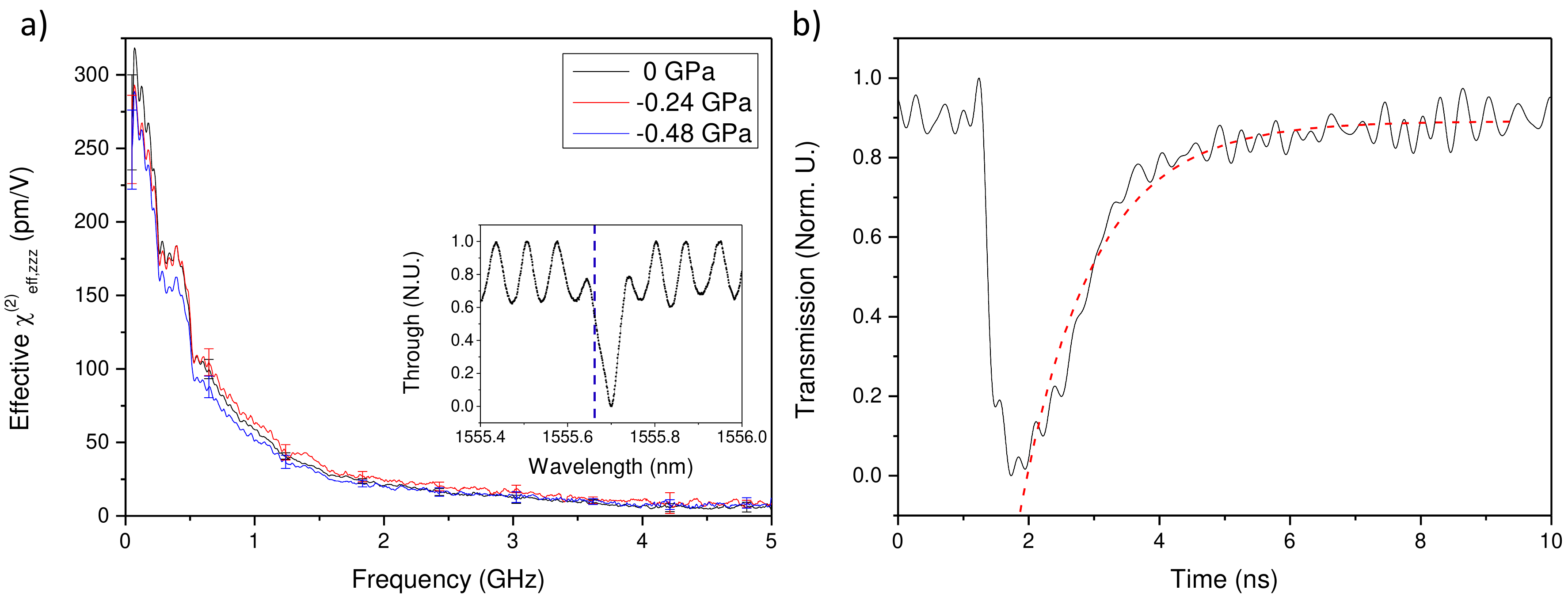}}
\caption{(a) Effective $\chi^{(2)}_{\textup{eff}}$ as a function of the electrical modulation frequency for three different stress levels in the waveguide. For clarity, the errors are reported only for certain points of the curves. The inset shows the working point of the electro-optic measurement. The black line denotes the Through transmission while the vertical blue dashed line represents the wavelength of the laser. (b) Measurement of the carrier lifetime. The black line is the Drop signal (in normalized units) in time while the red dotted line is an exponential fit of the rising edge.}
\label{fig:figure3}
\end{figure*} 
\noindent By monitoring the output signal at the reference photodiode (fig. \ref{fig:figure1}(c)), we tune the laser wavelength near the $-3\,dB$ point of one of the resonances (fig. \ref{fig:figure3} (a) inset), where the sensitivity to small refractive index variations is maximized. The quality factor $Q$ of this resonance is $\simeq 23200$. We then apply a sinusoidal voltage to the sample electrodes (fig. \ref{fig:figure1} (a)) using the VNA and a $32\,dB$ electrical amplifier. As a result of the bias modulation, the resonance oscillates back and forth with respect to the laser wavelength, inducing a periodic modulation of the transmitted optical signal at the Through port of the resonator, similarly to the operation of a conventional resonant ring modulator (RRM). Assuming that this signal modulation is due to the electro-optic effect caused by the strain induced second order nonlinearity, the oscillation amplitude $I_0$ is given by:
\begin{equation}
I_0 = \Bigl( \frac{\partial I_{\textup{out}}}{\partial\lambda_0}\Bigl) \frac{\bar{\lambda_0}L\chi^{(2)}_{\textup{eff},zzz}}{2L_{\textup{tot}}n_c^2}E_{DC}\label{eq:5}
\end{equation}
in which $\partial I_{\textup{out}}/\partial \lambda_0$ is the local slope of the spectrum, $\bar{\lambda_0}$ is one of the resonance wavelengths of the resonator, $L$ is the length of region where the voltage is applied, $L_{tot}$ is the resonator length, $n_c$ the refractive index of the core material of the waveguide and $E_{DC}$ the average electric field in the core (computed using FEM simulations and assumed completely polarized in the $z$ direction).
In Eq.\ref{eq:5}, the effective $\chi^{(2)}_{\textup{eff},zzz}$ is defined as:
\begin{equation}
\chi^{(2)}_{\textup{eff},zzz} = \frac{\int_{core}{\chi^{(2)}_{zzz}(z,x)n^2(z,x)|E_z(z,x)|^2dzdx}}{\int{n^2(z,x)|\textbf{E}(z,x)|^2dzdx}} \label{eq:4}
\end{equation}
where $\chi^{(2)}_{zzz}$ is the local strain induced $\chi^{(2)}$ in the material, $n$ the refractive index distribution and $\textbf{E}(z,x)$ the electric field of the optical mode. In Eq.\ref{eq:4}, we have assumed that the optical field is mainly polarized in the $z$ direction, so that off-diagonal tensor elements do not contribute to the integral. Eq.\ref{eq:5} can be derived from the fact that the Pockels induced effective index change $\Delta n_{\textup{eff}}=(\chi^{(2)}_{\textup{eff},zzz}E_{DC} n_g)/2n_c^2$ shifts the resonance frequency $\bar{\lambda_0}$ by a quantity $\Delta \lambda = (\Delta n_{\textup{eff}}L)/(L_{\textup{tot}}n_g)$. The VNA actually records $I_0$, from which, through Eq.\ref{eq:5}, we extract $\chi^{(2)}_{\textup{eff},zzz}.$
Figure \ref{fig:figure3}(a) shows the effective $\chi^{(2)}_{\textup{eff},zzz}$ values as a function of the driving bias frequency. 
The latter is swept from $50\,MHz$ to $5\,GHz$. As we can see, the value of $\chi^{(2)}_{\textup{eff},zzz}$ is maximum in the low frequency range, and decreases as the modulation frequency increases. We point out that the extracted $\chi^{(2)}_{\textup{eff},zzz}$ value close to DC regime is $\approx 270\,pm/V$, which is comparable to values reported in the literature for static electro-optic measurements using the same electric field and optical polarization directions \cite{chim2,puckett}. The cut-off frequency, i.e, the frequency at which the $\chi^{(2)}_{\textup{eff},zzz}$ halves, is $\nu_{c}=(0.50\pm0.01)\,GHz$, corresponding to a time constant $\tau = (0.55\pm0.01)\,ns$. The minimum value of $\chi^{(2)}_{\textup{eff},zzz}$ that can be detected is limited by the electrical noise floor of the VNA and of the photoreceiver. This corresponds to an effective $\chi^{(2)}_{\textup{eff},zzz}$ of $\approx 8\,pm/V$. From fig. \ref{fig:figure3}(a), one can see that this value is reached for  $\nu \ge 4.5\,GHz$. To exclude the possibility that the bandwidth is limited by the photon lifetime in the cavity, we modulated the optical input, and measured the signal at the Drop port with the laser wavelength tuned to the cavity resonance. The signal drops by $\approx -1.3\,dB$ from $50\,MHz$ to $5\,GHz$, showing that the cavity is far from the optical cut-off. This frequency response has been subtracted from the curves in fig.\ref{fig:figure3}(a).\\  
\noindent As shown in fig.\ref{fig:figure3}(a), we repeated the electro-optic measurements after applying stress levels to the waveguide of respectively $-0.24\,GPa$ and $-0.48\,GPa$. Very slight differences are observed with respect to the no-stress case. We also tested waveguides with smaller widths, such as $400\,nm$ and $800\,nm$, and found smaller transmission signal modulations compared to the $1600\,nm$ wide one.\\
\noindent These results clearly show that the modulation can not be attributed to a strain induced $\chi^{(2)}$, i.e., to the  linear electro-optic effect. In fact, if the latter were the cause for the observed effect,  one would expect the transmitted signal to follow voltage variations instantaneously up to optical frequencies. The signal modulation is rather related to a slower dispersion mechanism, with a characteristic time in the nanosecond scale. It has been recently shown that the plasma carrier dispersion effect can result in a linear electro-optic modulation if a charge layer is present at the $Si_3N_4-Si$ interface \cite{azadeh,rajat}. Therefore, we checked the effective carrier lifetime in our waveguide to verify its consistency with the cut-off frequency in the electro-optic experiments. We used a pump and probe scheme, in which an intense $ps$ laser pulse is coupled to the waveguide and the time dependent losses of a weaker probe beam are monitored. The short pump pulse generates free carriers due to two-photon absorption (TPA), these free carriers in turn attenuate the probe signal due to free-carrier absorption. After switching off the pump laser, the probe beam transmission slowly recovers due to free carrier recombination or diffusion away from the spot of the pump laser. The effective free carrier lifetime can then be extracted from the time-resolved measurement of the recovering probe beam transmission. The result is shown in fig. \ref{fig:figure3}(b). The sudden signal decrease is due to the pulse arrival and, consequently, to TPA carrier generation. The following slower signal recovery is due to the finite free carrier lifetime. From these data, we estimated a carrier lifetime of $\tau_c=(1.06\pm0.01)\,ns$. 
\noindent Being $\tau_c \simeq 1/\nu_c$, we conclude that the observed modulation can be attributed to plasma carrier dispersion. As further support to our conclusions, we applied static DC voltages to the electrodes of the racetrack resonator. We swept from $0\,V$ to $70\,V$ and measured the  spectral response at the Drop port as a function of the bias. We found that the maximum dropped signal monotonically decreases with voltage. At $70\,V$, the signal decreases by $1\,dB$ with respect to the zero bias condition. This behaviour confirms that the electro-optic modulation is actually induced by free carriers since it also introduces additional round trip lossess that would be unexpected in the case of strain induced Pockels effect. Furthermore, it is interesting to observe that the modulation frequency response is limited by the carrier lifetime. 
A plausible explanation of our observations is the accumulation and release of carriers at the $Si_3N_4-Si$ interfaces as a consequence of the applied voltages \cite{azadeh,zhang,alloatti}.\\
\noindent In addition, we set an upper limit to the strain induced $\chi^{(2)}_{\textup{eff},zzz}$ of $8\,pm/V$ at $-0.5\,GPa$ of applied stress. This value is about an order of magnitude lower than those extracted from DC measurements, as reported here and in the literature \cite{chim,chim2,puckett,pedro}. At DC, we believe $\chi^{(2)}$ induced modulation to be completely masked by free carrier effects. We remark here that our results do not exclude the presence of a strain induced $\chi^{(2)}$ in Silicon. Indeed, there exist proofs of Second Harmonic Generation (SHG) in strained Silicon that intrinsically can not have an explanation purely relying on free carrier  \cite{cazza,mher,Govorkov}. These experiments revealed a $\chi^{(2)}$ value of up to $40\,pm/V$ for $-1.2 \,GPa$ applied stress. Differences can be due to the fact that the $\chi^{(2)}$ tensor is dispersive, so its value at optical frequencies can be significantly different from the one measured at DC \cite{cazza}. In addition, it has recently been shown that electrostatic fields at the $Si_3N_4-Si$ interface can couple with the optical fields through $\chi^{(3)}$ nonlinearities, resulting in a Electric Field Second Harmonic Contribution (EFISH) \cite{mher}. This may have altered and consequently increased the extracted values of the $\chi^{(2)}$. 
 To summarize the results, in this work we demonstrate that a strong linear electro-optic effect is present in strained Silicon waveguides due to free carrier dispersion. A dedicated experimental set-up allows us to continuously tune the applied stress on a very same racetrack resonator. By performing high frequency measurements of the optical transmission under an AC electric field variation, we found that the electro-optic modulation vanishes as the modulation speed exceeds the free carrier lifetime. Thus, we evidenced a time response in the nanosecond range and, consequently, ruled out the potential strain induced $\chi^{(2)}$ as the origin of the modulation. We extract an upper limit of $8\,pm/V$ for the strain induced $\chi^{(2)}_{\textup{eff},zzz}$ in Silicon waveguides, which corresponds to our minimum detectable signal. This value is more than one order of magnitude lower than the one reported in the low frequency regime in the literature, which allows us to conclude that free carriers are responsible for the observed behavior. Larger stress or different stressing materials than $Si_3N_4$ are needed to definitely prove the presence of the electro-optic effect in strained Silicon waveguides. 

\noindent \textbf{Acknowledgment.} The authors acknowledge Mr.Saeed Sharif for the collaboration and useful discussions we had during the measurement sessions.
G. Pucker and M. Ghulinyan acknowledge the support of the staff of MNF during device fabrication.  This work has been supported by the SIQURO project financed by the Autonomous Provincia of Trento.

\end{document}